# A detailed interpretation of probability, and its link with quantum mechanics.

*Louis Vervoort (\*)*

*Université de Montréal, Montréal, Canada*

(Dated: November 29, 2010)

**Abstract**. In the following we revisit the frequency interpretation of probability of Richard von Mises, in order to bring the essential implicit notions in focus. Following von Mises, we argue that probability can only be defined for events that can be repeated in similar conditions, and that exhibit 'frequency stabilization'. The central idea of the present article is that the mentioned 'conditions' should be well-defined and 'partitioned'. More precisely, we will divide probabilistic systems into object, environment, and probing subsystem, and show that such partitioning allows to solve a wide variety of classic paradoxes of probability theory. As a corollary, we arrive at the surprising conclusion that at least one central idea of the orthodox interpretation of quantum mechanics is a direct consequence of the meaning of probability. More precisely, the idea that the "observer influences the quantum system" is obvious if one realizes that quantum systems are probabilistic systems; it holds for all probabilistic systems, whether quantum or classical.

**1. Introduction.**

The first and simplest axiomatic mathematical system for probability, published in 1933 by Andrei Kolmogorov, is reputed to cover all probabilistic systems of the natural, applied and social sciences. At the same time Kolmogorov's theory – the 'calculus of probabilistic or random systems' - does not contain the term 'probabilistic' or 'random', and does not provide any interpretation of the notion of probability, besides through the mathematical axioms it fulfills ([1], p. 100). For an intuitive understanding beyond mathematics of what probability 'is', one thus needs to resort to the ideas of other fathers of probability, Laplace, Fermat, Venn, von Mises, etc. Clearly, in developing his theory Kolomogorov[1] was much inspired by these interpretations, so much that one can read ([2], p. 2): "Kolmogorov himself followed a rather vague intuition of frequentist probability, an attitude most likely to be found shared today by probability theorists."

---

[1] In citing any usual number of contributors to probability theory, one does not escape from the feeling to do injustice to a long list of excellent mathematicians. At the time of Kolmogorov, substantial contributions to modern probability calculus had already been made - for instance – by Lebesgue, Borel, Bernstein, Khinchine, Markov, Liapunov,…

We indeed believe that in the broader physics community (and in the natural sciences in general) the most popular interpretation of probability is the frequency model – especially the limiting frequency version due to physicist and mathematician Richard von Mises [3,4]. However, in philosophy and the foundations of quantum mechanics other interpretations, in particular the subjective interpretation[2] associating probability and 'degree of belief', are increasingly popular (for references on the different interpretations of probability see [2,5,6]; more on the subjective interpretation in quantum mechanics in, e.g., [7]). At any rate, we believe that anyone who has tried to construct a precise definition of probability soon realizes that the topic is indeed surprisingly subtle. At the same time such a definition is highly desirable, for several reasons. First, suffices to have a look at the reference works of von Mises (or almost any reference text on the calculus) to realize that without a precise idea of what probability is, any somewhat subtle problem of calculus is likely to be treated in a wrong manner (von Mises provides many tens and probably hundreds of such flawed calculations by even skilled mathematicians). No surprise, it has been said that "In no other branch of mathematics is it so easy to make mistakes as in probability theory" ([8], p. 4). Next, for foundational reasons it is even more important to have a clear idea of the concepts, as we will illustrate here by exposing a link between the interpretation of probability and quantum mechanics. The interpretation of probability would thus appear to be a wonderful topic at the interface of mathematics, physics, and philosophy.

In our view, a satisfactory definition should be overarching and apply to both chance games, usually treated by the classical interpretation, and natural probabilistic phenomena, usually treated by the frequency interpretation. In the present article we propose such an overarching definition or interpretation; it is the main topic of this article. To our surprise, we discovered that a precise formulation of the definition immediately suggests an intimate link with quantum mechanics: one of the fundamental ideas of quantum theory, namely that "measurement determines the quantum system" [9], appears to be nothing more than a feature of *any* probabilistic system, quantum or classical. If that is true the link between quantum and classical (probabilistic) systems is stronger than usually thought; this result would be an immediate consequence of a precise definition of probability, going a little further than 'relative frequency for n → ∞'.

---

[2] We believe that the true nature of probability is essentially to be found in the real world around us, more so than in our heads – indeed, we will provide arguments it can be defined in an objective manner.

In more detail, we will propose in the following a model that can be seen as a variant of von Mises' interpretation [3,4], differing, however, in two essential respects. First, our model simplifies von Mises' theory. In our view von Mises' work (especially his [4]) is essential in that it offers the most rigorous interpretational analysis of all mathematical aspects of probability calculus (for critiques of this position, see Section 2). As is well known, von Mises introduces, to that end, the concept of 'collective', and proposes a detailed calculus for it – a calculus that strikes however by its complexity, and that very probably no-one would really consider using nowadays on a regular basis (see e.g. von Mises' treatment of de Méré's problem, [3] p. 58ff). We will not make use of the concept of collective; in particular we believe that it is not necessary to resort to the notion of 'invariance under place selection' to characterize randomness (see next Sections, and also [6] p. 112). Our attitude is pragmatic: few people challenge the idea that the mathematics of probability theory is fully contained in Kolmogorov's calculus. What *is* however highly controversial is following question, intimately linked to the meaning of probability: *what is the exact subject matter of probability theory – to what exactly to apply it ?* We will push the answer further than von Mises' classical answer: according to him probability theory treats "mass phenomena and repetitive events" [3], characterized by a set of attributes that form the attribute space Ω. (Anticipating, we will arrive at the conclusion that probability theory applies to a special type of random events, which we will term 'p-random'.) As stated, above question is intimately related to the detailed interpretation of probability. The definition we will propose in Section 3 captures, we believe, the essence of von Mises' theory, but in a substantially simpler form.

Besides a simplification, our model allows a clarification of the frequentist interpretation. The main claim of the present article is that one substantially gains in *partitioning* probabilistic systems into subsystems, namely test object, initiating, and probing subsystem. We will argue that this partitioning allows to solve a wide variety of classic problems linked to the notion of probability. That is why we will conclude, in a slogan, that probabilistic systems are best considered as being *composed*.

The present text is organized as follows. In Section 2 we focus our attention on a simple random system, namely a die, and see how the frequentist model should be applied here. This allows us to introduce two key ideas of the present text, namely that it is helpful, if not necessary, to partition probabilistic systems in subsystems, and that the essential feature of such systems is the property we will term 'frequency stabilization' (a term that occasionally appears in von Mises' work, and that in our opinion best captures the property in question). We will in this Section also succinctly review the most frequently encountered criticisms of von Mises' interpretation. In Section 3 we argue that a confusion exists concerning the attribute 'random': not all randomness is the same. We will generalize the ideas of Section 2, and formulate definitions for randomness and probability. In Section 4 we will argue that the model allows to gain insight in classic paradoxes of probability theory, such as Bertrand's paradox. Finally, the link between classical and quantum systems will follow as an immediate corollary of our definition.

Many of these topics can obviously not be treated here in full detail. We can often only provide a first entry into these matters, and are well aware that many questions will remain unanswered. We hope however that the variety of problems our model allows to tackle lends it credit as a candidate that deserves further attention.

**2. A simple random system: a die. Introductory ideas: frequency stabilization and partitioning.**

What is a probabilistic, or random, event ? When does it make sense to speak about probability ? These are much debated questions in philosophy and foundational research, and most texts conclude that no consensus has been reached on these matters [2,5,6]. Observe that they can be asked within *any* of the existing (objective) interpretations. Let us, *en passant*, note following observation we find particularly puzzling: how is it possible that such very different systems as dice, playing cards, quantum systems, the distribution of human lengths, roulette wheels, population features, diffusion phenomena, measurement errors, etc. are all governed (described) by *one and the same* (rather simple) mathematical theory, namely probability calculus ? How is it possible they share a common ontological element ? – some might be inclined to ask. In view of the multiplicity of probabilistic systems, it may not be surprising that several interpretations of probability exist; on the other hand the uniqueness of the axiomatic system seems to cry out for one overarching interpretation.

In order to get a feel for above questions, and to be able to introduce a few essential notions, let us have a look at a typical probabilistic system, namely a die. When or why is a die 'probabilistic', or 'random' (or rather the throwing events, or outcomes) ? Simply stating in non-anthropocentric terms what a die throw is, seems to immediately bring to the fore a few important notions. In physical terms a 'die throwing event' or 'die throw' consists of the 3-dimensional movement of a (regular) die, that is 1) characterized by the time evolution of its center of mass and its three Euler angles, 2) *caused* or *initiated* by an 'initiating system' (e.g. a randomizing and throwing hand), and 3) *probed* by an 'observation system' (e.g. a table) that allows to observe (in general 'measure') an outcome or result R (one up, two up, etc.). It is easy to realize that if we want to use the die 'as it should', i.e. *if we want to be able to observe or measure a probability* for the different results of the throws, we have to repeat the throws not only by using the same die (or similar regular dies), but also by using the same 'boundary conditions'. These boundary conditions, then, are related to our throwing, the table, and the environment in general. *Irregular conditions in any one of these three elements may alter the probability distribution*. We can for instance not substantially alter our hand movement, e.g. by putting the die systematically ace up on the table. Nor can we put glue on the table, position it close to our hand, and gently overturn the die on the table while always starting ace up – six up could happen substantially more often than in a ratio of $1/6^{th}$. Nor can we do some experiments in honey in stead of air: again one can imagine situations in which the probabilities of the outcomes are altered. As will be seen further, it will make sense to isolate the mentioned elements, and to consider our *random event* as involving a *random system* containing three subsystems, namely 1) the (random) *test object* itself (the die), 2) the *initiating system* (the throwing hand), and 3) the *probing* or *observation system* (the table, and let's include the human eye). Just as one can associate subsystems to the random event, one can associate (composed) *conditions* to it (conditions under which the event occurs), in particular initiating and probing conditions.

Let us immediately note that instead of 'initiating system' and 'probing system', it can be more appropriate to speak of 'environment', namely in the case of spontaneous or 'natural' probabilistic events (versus 'artificial' ones, as outcomes of chance games, which are created by human intervention). Indeed, many random events occur spontaneously, without any known cause. A spontaneously disintegrating nucleus has probabilistic properties (e.g. its half-life, i.e. the time after which it disintegrates with a probability = 1/2) that are, on usual interpretations, not 'initiated', as in our die throwing, by a physical system or a person. Neither is it 'probed', except when subject to experiments in a laboratory. But the nucleus disintegrates spontaneously, according to a given well-defined probabilistic pattern, *only in a well-defined environment*. Changing the environment, e.g. by irradiating the nuclei or by strongly heating them, may very well change the probability distribution in question. In other words, if we want to determine the half-life of the nucleus, i.e. the probability of disintegration, we have to put it in well-defined ('initializing') conditions of temperature, pressure, etc., and measure its properties in well-defined ('probing') conditions, that scrupulously imitate its natural environment – so also here the initial and final conditions, or environment, re-appear.

By the above partitioning in subsystems, we have just rendered the concept of 'conditions' explicit – a concept that is the starting point of, e.g., the reference work of Gnedenko on probability calculus [10][3]. Gnedenko states (p. 21):

> "On the basis of observation and experiment science arrives at the formulation of the natural laws that govern the phenomena it studies. The simplest and most widely used scheme of such laws is the following: *Whenever a certain set of conditions C is realized, the event A occurs.*"

And a little further ([10] p. 21):

> "An event that may or may not occur when the set of conditions C is realized, is called *random*."

We will extensively come back to this partitioning of the conditions and of the probabilistic system in the next Sections. We will argue that it may form the basis of a unified treatment of probabilistic systems. For chance games, treated by the classical interpretation, the initiating subsystem is for instance simply our randomizing hand; for spontaneous, natural phenomena, usually treated by the frequency interpretation, the initiating subsystem is simply part of, or is, the environment.

---

[3] We found this often cited work most helpful in our study. Indeed, Boris Gnedenko, a pupil of Kolmogorov, did not only substantially contribute to probability calculus (especially in applied statistics and elaborations of the Central Limit Theorem), he was also, as Kolmogorov, highly interested in the foundational issues – as his book manifestly reflects.

For the well-known physical system of a thrown die, that is now well-defined, it seems not difficult to deduce the conditions for 'randomness' (we will see further that the condition proposed above by Gnedenko is not sufficient). Indeed, in order that the die throws are 'probabilistic' or 'random' or 'stochastic', it is a necessary condition that the *relative frequencies of the results* $R_j$ (j = 1,…, 6), namely the ratios { (the number of events that have result $R_j$) / n }, where n is the number of trials, *converge towards a constant number* when n grows (1/6 for a true die). If there would be no such '*frequency stabilization*', one cannot speak of probability. Indeed, if for instance the die sublimates during the experiment in an asymmetric manner (thus gradually loosing weight on one side), or if we accompany the die's movement in a structured manner that does not lead to frequency stabilization, we simply cannot attribute a probability to the outcomes. In sum, if there is no frequency stabilization, the die throwing is not probabilistic; on the other hand, if frequency stabilization occurs, the probability of the result $R_j$ is given by $P(R_j)$ = (# events that have result $R_j$) / n for large n. This is, or this is in full agreement with, the frequency interpretation of von Mises and others.

Note that, as formulated above, random systems include deterministic systems. Indeed, in the latter case frequency stabilization will also occur. If we move the die by placing it systematically – deterministically – ace up on the table, the six $P(R_j)$ will trivially converge to a value (namely to 1, 0, 0, 0, 0, 0). We could exclude deterministic cases from our definition of random events / systems by adding that, besides exhibiting frequency stabilization, the events / systems should be unpredictable. Thus 'genuinely' random events show frequency stabilization *and* are unpredictable (or one could replace the latter by another predicate that qualifies the 'unstructured', disordered nature of randomness).

Before generalizing these ideas, let us have a look at some criticisms of the frequency interpretation, as they appear in some foundational works. It is well known that von Mises' frequentist theory suffers from minor shortcomings in strict mathematical rigor (see e.g. [2] Chap. VI). Von Mises uses in his theory the concept of *limiting* relative frequency (for the limit n → ∞)[4], which is considered problematic by some. He also characterizes randomness by using the concept of - again infinite - 'collectives' (a collective is the in principle infinite series of trials that intervenes in the definition of probability, see former footnote). But this mathematical shortcoming does not prejudice in this case, we believe, the overall value of the interpretation - and the many contributions to the calculus – that were offered by von Mises. The most relevant critique of the mathematics of infinite collectives comes arguably from Jean Ville [11], who concludes his study however with the essential remark that an expression as "the frequencies have a tendency to group around a certain value" seems impossible to avoid, even if mathematically not entirely rigorous (cited in [2], p. 197). We would think that mathematical rigor is not the first virtue of theories when they aim at explaining real phenomena, as von Mises often emphasized. The concept of infinity is ubiquitous in theoretical physics (e.g. when taking integrals over infinite ranges), even if it is obviously an idealization – a useful one when the formulas correctly predict reality to the desired precision (moreover, it seems most of the time not difficult to explain in concrete cases *why* the limiting case for 'n $\rightarrow \infty$' works – i.e. is a good approximation of reality).

---

[4] Probability is defined as $\lim_{n \to \infty} n(R_j)/n$, where $n(R_j)$ is the number of events or trials that have result $R_j$ in the total series of n trials, as in above example.

In view of the results of von Mises, Ville, and others, we believe that the limit n $\rightarrow \infty$ does *not* pose a problem, but reflects to the contrary a fact of nature (see also C6 below). Indeed, at the very basis of the frequency interpretation lies the scientific hypothesis that if certain well-defined experiments (trials) *would* be repeated an infinity of times, certain ratio's (involving outcomes of these trials) would converge to a limit. Experiment has shown over and over again that *long* series (obviously not infinite ones) *do* show such frequency stabilization, corroborating the initial assumption as strongly as any scientific hypothesis can be confirmed (scientific hypotheses are never confirmed with absolute certainty, neither with absolute precision). At any rate, we will in the definitions we propose accommodate certain criticisms, since we will *not* define frequency stabilization in terms of convergence in the limit n $\rightarrow \infty$, but 'for n large enough for the desired precision' (more on this in C6 below).

More important, we believe, is to understand the essential reason for von Mises' introduction of the concept of 'collective'. The latter allows, according to its spiritual father, to define what a 'random' event is; only when randomness is defined, probability can be defined (the definition of probability uses the definition of randomness). Even if, seen from a historic perspective, von Mises' calculus of collectives did not have much success due to its complexity, we will corroborate in the next Section his original thesis that probability can only be defined for random systems, and that randomness is logically prior to probability, in the sense just stated. We believe however that the notion of 'collective' is not needed to define randomness, as shown in the next Section.

As last critique, let us mention Fine [5], who derives a theorem (p. 93) that is interpreted by its author as showing that frequency stabilization is nothing objective, but "the outcome of our approach to data". However, closer inspection shows that Fine's argument only applies to complex mathematical sequences (of 0's and 1's) that can be generated by computer programs; and if a sequence can be generated by a short computer program, *it is by definition not random, but deterministic* (it may of course look random). An essential point of real series of coin tosses is that they cannot be predicted by numerical algorithms… *and* that they show frequency stabilization, as can easily be shown by experimentation. From this perspective, mathematical randomness or complexity, as discussed in relation to the complexity of number strings, has little or nothing to do with physical randomness[5].

**3. Generalization, definitions.**

Before proposing our definitions, which will essentially be straightforward generalizations of what we learned in the preceding Section, let us first sketch the boundary conditions, and the essential problems the model has to tackle.

While studying in some detail our die, we found that the essential feature qualifying the throws as random, or probabilistic, is frequency stabilization. It seems just a natural interpretation of the frequency model to put this characteristic at the centre of the theory, as we will do further on. That is indeed what von Mises did ([3], p. 12). He writes:

> "It is essential for the theory of probability that experience has shown that in the game of dice, as in all the other mass phenomena which we have mentioned, the relative frequencies of certain attributes become more and more stable as the number of observations is increased."

But here an obvious, but neglected, question pops up: are all random systems really random ? At this stage of analysis, the answer seems not difficult: if frequency stabilization is indeed the essential feature of random systems and events, then it is clear that *not* all so-called random systems are random in the sense of probabilistic or stochastic. *Frequency stabilization is by no means guaranteed for an arbitrary disordered event, system or parameter*. The duration of my Sunday walks looks random, but does this parameter exhibit frequency stabilization ? No, because 1) there are systematic variations during certain periods, e.g. during the 'lazy years', 2) with age the walks will systematically tend to become shorter, and 3) at some point they will unfortunately disappear altogether. Or consider a mountain torrent, and define a parameter characterizing its flow, such as the (average) height of its waves or its rate of flow through a given surface. Such a torrent may show such erratic behavior over time, e.g. due to periods of rain or drought, that it seems impossible that any parameter can be defined of which the frequency would stabilize. Atmospheric temperatures in any given city look random, they may even look statistically random, but they probably don't show frequency stabilization over long periods because of climate changes[6]. What is even less likely to exhibit frequency-stabilized features, is 1) any 'random' system that is subject to human actions, e.g. the chaos on my table, or the random-looking features of a city-plan, etc., and 2) any 'composite' random-looking system, like a car and the house in front of it, or a tree, a stone underneath, and a cloud above. In other words, frequency stabilization happens often[7], but certainly not always. *Not all randomness is the randomness that is the object of probability calculus*.

---

[5] Except if the number strings are generated by a real physical number generator, which is almost never the case.

Above observation justifies the introduction of the concept of 'p-randomness', to be contrasted with 'randomness' and 'non-p-randomness': p-randomness, or probabilistic or structured randomness, is randomness (chance) that is characterized by frequency stabilization (at infinity, or if one prefers 'for large n'). Before one applies probability calculus to an event, one conjectures, or has evidence, that it is p-random - a feature that can only be confirmed by experiment. Again, from this point of view deterministic systems or events are closer to statistically random systems than non-statistically random systems are: both former systems show frequency stabilization. Note that a technically more familiar way to characterize a parameter or variable R as 'p-random', is to say that it has a 'probability density function', a parameter often symbolized by 'ρ' in physics (ρ = ρ(R)).

---

[6] This may however be seen as a limiting case, in which frequency stabilization occurs 'for large n', even if not 'at infinity': statistics seems still applicable here.
[7] A particularly beautiful mathematical argument for the ubiquity of probabilistic randomness is provided by the uncrowned queen of probability calculus, namely the Central Limit Theorem.

Surprisingly, and to the best of our knowledge, in texts on this topic one does not distinguish between unstructured randomness and the structured or probabilistic randomness that is the object of probability calculus. The distinction is implicit in Gnedenko's [10]. On p. 22 he writes:

"However, a wide range of phenomena exists for which, whenever the set of conditions C is realized *repeatedly*, the proportion of occurrences of the event A only seldom deviates significantly from some average value, and this number can thus serve as a characteristic index of the *mass phenomenon* (the repeated realization of the set of conditions C) with respect to the event A."

And probabilistic randomness is defined more explicitly on p. 156:

"A random variable is a variable quantity whose values depend on chance and for which there exists a distribution function".

Let us now ask our initial question again – or better cite once more Gnedenko, who often formulates precise foundational questions (p. 23, italics by him)[8]:

"*Under what conditions does the quantitative estimate of the probability of a random event A by means of a definite number P(A) – called the mathematical probability of the event A – have an objective meaning, and what is that meaning ?*"

Before trying to answer this question by our model, let us emphasize again that the probabilistic nature of a phenomenon can - in general - only be determined by experiment. Indeed, the only way to prove the probabilistic nature of systems or events is to do experiments, and to show that frequency stabilization occurs (which is equivalent, in technical terms, to proving that a density function ρ exists). As we saw in the preceding Section, it seems essential to clarify the conditions under which such experiments need to be done. The experiments involve repeated tests, in which the object under study (e.g. the die) should be subjected to repeatable, *'similar'* or *'identical'* conditions before being probed. It is however a well-known and unsolved problem of the foundations of probability to identify what 'similar' / 'identical' exactly means. How similar is similar[9] ? We all know by experience that we do not need a lot of precautions when shaking a die so as to launch it in a 'regular' manner – leading to the probability distribution of a true die. In other words, it seems that in general we have a good idea what these 'similar conditions' are; but that is not always the case. At this point the idea of partitioning we introduced in the preceding Section proves helful. Indeed, similar tests on similar systems means, in general: similar initiating system (or environment), similar test object, and similar probing system. These *three* subsystems should 'act', in the repeated test sequence, in *similar (enough) ways* in order that a probability can be defined. Alterations or divergences in any of these three subsystems can lead to different probabilities (probability distributions) – the latter may not even be *defined*, i. e. *existing*. (Remember the sublimating die, or the inadequate randomization etc., which can lead to an undefined probability distribution of the die throws.) But again, how similar should these subsystems, or the events they generate, be ?

---

[8] Gnedenko does not offer an explicit answer.

After having indicated the essential problems, we now propose our model, following the frequentist methodology.

DEF1. A system or event possesses the property of frequency stabilization IFF

(i) it is possible to repeat n 'identical' (or 'similar') experiments (with n a number sufficiently large for the desired precision) on the 'identical' object of the system by applying n times 'identical' initial and final actions or conditions on the object (realized by the initiating and probing subsystems, or more generally, by the environment); and

(ii) in this experimental series the relative frequencies of the results $R_j$ (j = 1,…, J) of the corresponding events (detected by the probing subsystem) converge towards a constant number when n grows; more precisely, the ratios { (# events in the series that have result $R_j$) / n } converge towards a constant number when n grows, for all j.

---

[9] Also Gnedenko is unclear here, and states e.g. ([10] p. 19): "By a mass phenomenon we mean one that takes place for aggregates of a large number of objects that have equal or almost equal status […]".

DEF2: Only for an event (or system, or outcome, or variable) that shows frequency stabilization according to DEF1, the probability of the result or outcome $R_j$ (j = 1,…, J) of the event is defined, and given by (for all j):

$P(R_j)$ = (# events that have result $R_j$) / n,   for n large enough for the desired precision.

We conjecture that above definitions allow to characterize all (objective) probabilistic phenomena; i.e. that they cover all events or systems described by probability calculus. To some believers of the frequency interpretation, this claim may seem acceptable or even obvious. Others however may find that a heavy burden of providing arguments awaits us. Clearly, we cannot present a full review of all probabilistic systems and show in detail how the definitions apply in all these cases. But under C1 – C6 we will provide those arguments and completing observations that seem the most important to us in rendering above conjecture defendable. In the next Section we will focus on the benefits of the model, i.e. address the problems and paradoxes of probability theory that the model allows to solve.

C1. It is well-known that the frequencies or ratios defining probability in a model like the above fulfill the axioms of Kolmogorov's probability calculus.

C2. If one reduces the definitions to their essence, they state that probability *only* exists for systems or events that can be subjected to *massively repeated physical tests, occurring under well-defined conditions*; and that in that case probability is given by a simple ratio or 'frequency'. The numerical precision of that ratio grows when the number of experimental repetitions grows. A particularity of the model is that it partitions the 'experimental conditions'. One advantage of our phrasing is that the definitions are immediately applicable to both 'artificial' systems (chance games, typically) and 'natural' systems (behaving in a p-random manner without human intervention); in other words our phrasing allows to see how the same concept applies to such widely different events as a die thrown by a human, and a molecule moving in a gas. Indeed, in the case of chance games the initializing and probing subsystems are easily identified as truly separate systems acted upon, or made, by humans (see C5). In the case of natural probabilistic phenomena (quantum phenomena, diffusion phenomena, growth phenomena, etc.) the initializing and probing systems coincide with – are - the environment[10].

Note that DEF1 of p-randomness is stated in terms of 'possible' repetitions, or of 'possible' trials, not necessarily actual ones. Natural, spontaneous random events are normally not initiated nor measured by humans in repeated experiments. Still, according to the model we advocate, they only 'have' a probability if they occur, or can occur, on a massive scale as can be imitated in the lab, or studied under repeatable conditions. Also for such systems probability only exists *if massively repeated tests are possible, and if frequency stabilization occurs*. The probability of such natural events is equal to that revealed by artificial experiments imitating the environment (or by calculations using experimental data, and representing such experiments): this is, we believe, nothing more than the scientific paradigm.

C3. Most p-random phenomena or events are characterized by outcomes that are (described by) *continuous* variables, rather than discrete ones. In other words, R will range over an interval of the real numbers, rather than taking discrete values $R_j$, j=1,…, J as in die throws, coin tosses and other chance games. One could observe that these objects used in games are constructed on purpose by humans in order to 'produce' a limited number of discrete outcomes – in contrast to natural systems. The latter, like diffusion phenomena, gas kinetics, biological growth phenomena, etc. have outcomes that are continuous, or quasi-continuous. (A notorious exception are quantum systems, notably when they are carefully 'prepared'.) In the case of a continuous p-random variable x (we used R before), one defines in probability theory the probability density function $\rho(x)$, the meaning of which is however defined via the concept of probability. Indeed, in the continuous case $P(R_j) = \rho(R_j).dR$, defining $\rho$ via P as characterized in DEF2. One can thus, for the present purpose, safely treat probability density functions and probability on the same foot. A model that interprets probability for discrete variables, also does the job for the ubiquitous continuous variables. (Also note in this context that one can formally describe discrete cases by a density function $\rho(R) = \Sigma_j P_j.\delta(R-R_j)$, where $\delta(-)$ is the Dirac delta-function: the discrete case is a special case of the more general continuous case.)

---

[10] When these phenomena are studied in the laboratory the attribute 'natural' loses some of its pertinence; reintroduction of the concepts of initializing and probing subsystems seems then appropriate again, in order to characterize the laboratory experiment in the detail that is necessary to define the probability of the studied event.

C4. DEF1 of p-randomness relies on the notion of 'identical' or 'similar' conditions and objects. It may thus, as already stated, look suspicious in the eyes of a philosopher. But we will now argue that the definition is sound in the sense that it allows to define p-randomness and probability in a fully objective manner. First, notice that it seems impossible to define 'similar' or 'identical' in a more explicit manner. One could try in following (partly circular) way: "to the best of common knowledge, or of expert knowledge, one should repeat the 'same' event in a manner that allows to speak about the 'same' event, and that generates frequency stabilization." But does this help ? We believe however there is no real problem. Indeed, as defined, *frequency stabilization can be tested for by independent observers; and these observers can come to the same conclusions*. The conditions for doing the 'frequency stabilization test' can be communicated for any particular system: "do such-and-such ('identical') initial and final actions on such-and-such ('identical') systems - and the probabilities $P_j$ will emerge. I found frequency stabilization and the probabilities $P_j$, so if you do the experiment in the 'right, identical' conditions, you should find them to." It would seem that the problematic term 'identical / similar' of probability theory can thus be defined in an operationally fully consistent manner.

In sum, the above shows that one could, or should, speak of 'p-identical', or rather 'ρ-identical', events / systems, where ρ is the probability density of the outcomes of the events under study: to test whether a system is probabilistic, and to identify the probability of the corresponding events, one needs to perform a series of experiments on ρ-identical systems (including object, environment, initializing / probing subsystem) – systems that lead to the probability distribution ρ(R) of the results R of the events 'defined' or 'generated' by the system. Upon this view, then, *'identical' is 'identical' insofar a probability distribution ρ emerges*. It will also be clear that in an analogous way one could meaningfully define a ρ-system = ρ-{object, environment, initializing /probing subsystem}. We therefore believe that in the context of probability theory, one should, in principle, speak of ρ-systems, or ρ-identical systems.

C5. Does the frequentist model cover the classical interpretation of probability, traditionally used to tackle chance games, urn pulling and the like ? Von Mises ([3], p. 66 ff.) and many modern texts on probability calculus come to this conclusion. We will here only present, concisely, what we believe to be the essential arguments for an affirmative answer to above question. Notice, first, that our definitions can at least in principle be applied to such chance games. The initializing subsystem is most of the time a 'randomizing hand' (tossing a coin or a die, pulling a card or a colored ball, etc.); the probing subsystem is often simply a table (plus a human eye).

According to the famous expression of Pierre-Simon Laplace, for calculating a probability of a certain outcome, one should consider "events of the same kind" one is "equally undecided about"[11]; within this set, the probability of an outcome is the ratio of "favorable cases to all possible cases". In all reference books on probability calculus, the latter definition is the basis of probability calculations in chance games and urn pulling. Notice now that dice, card decks, urns containing balls, roulette wheels, etc. are constructed so that they can be used to produce *equiprobable* (and mutually exclusive and discrete) *basic outcomes*, i.e. *having all $P_j$ equal, and given by 1/J* (J = the number of basic[12] outcomes). Equiprobability is at any rate the assumption one starts from for making mathematical predictions, and for playing and betting; and indeed Laplace's "events of the same kind one is equally undecided about" would now be termed equiprobable events. Along the lines exposed above, a chance game can thus be seen to correspond to an (artificially constructed) ρ-system with $\rho(R) = \Sigma_j (1/J) \delta(R-R_j)$.

It is at this point simple to show that in the special case of equiprobable and mutually exclusive events, the frequentist and classical interpretation lead to the same numerical values of probability. Indeed, within the frequentist model, $P(R_j) = \dfrac{n.\dfrac{1}{J}}{n} = 1/J$ (the numerator n / J = the number of $R_j$-events among n (>>J) exclusive and equiprobable events each having a probability 1/J). Thus the result, 1/J, is equal to the prediction given by Laplace's formula (1 favorable case over J possible cases). Let us immediately note, however, that Laplace's formulation is not superfluous. It allows, for the special case of equiprobable events, for calculation: 'favorable cases' and 'all possible cases' can indeed conveniently be calculated by the mathematical branch of combinatorics (a theory of counting, initiated by the fathers of probability theory).

---

[11] The events thus fulfill the 'principle of indifference' introduced by Keynes.
[12] The 'basic events' or 'basic outcomes' of coin tossing are: {heads, tails}, of die throwing: {0, 1, …, 6}, etc. The probability of 'non-basic', or rather composed, events (two consecutive heads, etc.) can be calculated by using the theory of probability calculus and combinatorics.

In sum, it appears that the classical interpretation is only applicable to a small subset of all probabilistic events and systems, namely those characterized by discrete and equiprobable basic outcomes. For this subset Laplace's interpretation can be seen as a *formula*, indeed handy for calculations, *rather than an interpretation of probability*. After calculation *the only way to verify the prediction is by testing for frequency stabilization*. It is among others for this reason we believe the latter property is the natural candidate for a basis of an overarching interpretation.

C6. We have defined frequency stabilization and probability by means of the notion of 'convergence' of a certain ratio when n, the number of trials or repetitions of an experiment, 'grows'. It is clear that this phrasing is close to the original definition by von Mises of probability as $P(R_j) = \lim_{n \to \infty} n(R_j)/n$. However, our phrasing "$P(R_j) = n(R_j)/n$ for a number of trials n that is large enough for the desired precision" avoids the notion of infinity; it may therefore avoid problems of mathematical rigor (see discussion in Section 2). Note that from a pragmatic point of view, our definition allows to derive, if one would use it to experimentally determine a probability, numbers that are equal to those identified by von Mises' definition to any desired precision. At least operationally there is no difference in the definitions: they lead to the same results. Von Mises' definition may, however, be more satisfactory to our metaphysical aspirations: one could say that probability indeed 'is' limiting relative frequency 'at infinity'. An 'ideal' number that one can seldom calculate with infinite precision (except, for instance, if one can use combinatorics), but that one can always measure with a precision that grows when n grows.

These notes conclude the basic description of our model. Needless to say, they are a first introduction into the model; we are well aware that questions will remain unanswered. The only way to validate and strengthen a model is to show that it applies to all non-controversial cases, *and* that it solves problems left open by other models. The more it can do so, the more it convinces us that it captures the meaning of probability - beyond the pure mathematics, which is covered by Kolmogorov's calculus. Concerning the 'non-controversial cases', we could have shown in more detail how the model applies to other typical systems, e.g. gas molecules (this is one of the model systems von Mises investigates). We believe that the reader will however have no problem in applying the above model also to this case (see also von Mises' [3] p. 20); again, the key for such an application is to recognize that the initializing system here is simply the environment. Since the above model is an elaboration of von Mises' interpretation, it should be able to tackle all probabilistic systems that the latter can tackle. However it is much simpler: we do not need the concept of collective, nor its calculus; our calculus is Kolmogorov's.

**4. Further results, and link with quantum mechanics.**

Let us now derive some further results, and in particular see which problems and paradoxes the model allows to solve. The point that will interest here in particular is R3, where we will investigate what above model says about quantum mechanics.

R1. As is well known, according to frequency interpretations it makes no sense to talk about the probability of an event that cannot be repeated. In condition (i) of our definition of p-randomness, repeatability is an explicit requirement. Therefore, it makes no sense to talk about the 'probability' of Hitler starting a world-war and the like: no experiments can be repeated here, and even less experiments in well-defined conditions. Popper's propensity interpretation of probability was an attempt to accommodate such probabilities of single events – quantum events were his inspiration, like the disintegration of one atom ([6] Ch. 6). But again, any quantum property of any single quantum system can only be attributed a probability if the property can be measured on an ensemble of such systems – exactly as in the case of macroscopic systems. Measurement (verification) of a probability is *always* done on an ensemble of similar systems and events, whether quantum or classical.

According to our model, probability is *not* a property of an object on its own; it is a property of certain systems under certain repeatable human actions; or, more generally, *of certain systems in dynamical evolution in well-defined and massively repeatable conditions, in particular initializing and probing conditions*. Following slogan captures a part of this idea: *probability is a property of certain composed systems*. Similarly, probability is not a property of an event *simpliciter*, but of an event in well-defined conditions – of certain 'composed' events. Another roughly equivalent way to summarize these ideas would be, it seems, to attribute probability to *experiments*. The advantage of the term 'experiment' is that it only applies to 'composed events' for which the experimental conditions are defined in a scientifically satisfying manner – exactly as we believe is the case for a legitimate use of 'probability'. Note that a scientific experiment is also, by definition, repeatable – Hitler's 'experiment' is not. In sum, in our view probability is a property of composed events or composed systems, or of experiments, and not of events or objects *simpliciter*.

R2. Let us have a look at the famous paradox of Bertrand, for instance as discussed in van Fraassen ([12] p. 303), who proposes following version of it:

"A precision tool factory produces iron cubes with edge length ≤ 2 cm. What is the probability that a cube has length ≤ 1 cm, given that it was produced by that factory."

In his Chapter 12, van Fraassen shows that the 'principle of indifference' of the classical interpretation of probability is too vague to be tenable, and he offers arguments, one of them being Bertrand's paradox.

What does our frequentist model tell us about this paradox ? As we argued under C5, the principle of indifference is indeed only valid for a small set of probabilistic systems, notably for the chance games of classical probability, more precisely for systems that produce equiprobable basic events (and thus flat distribution densities). Only in that case we are justified to be 'equally undecided' about a particular basic outcome to be produced. Most of the time however we do not a priori know the probability distribution of the outcomes. Only in simple chance games we know a priori it is flat, because the systems were constructed to that end; but in most systems some outcomes are more probable than others. And indeed, as is well known by statisticians and engineers, the precision fabrication processes of van Fraassen's version usually do lead to a characteristic non-flat distribution of the side lengths, namely the ubiquitous Gaussian probability density, centered on the average side length. However, due to so-called 'systematic deviations', the distributions could be non-Gaussian. There is no 'principle of indifference' that could help us to identify the distribution a priori – we need empirical information (van Fraassen comes to the same conclusion). In this light, it is clear that Bertrand's problem, in above version or any other, is badly posed (with the given information, an infinity of answers is possible – one for each distribution function)[13].

This conclusion is an immediate consequence of our model. DEF1+2 directly imply that the problem, i.e. the sought probability, is only defined *if the factory produces p-random cubes*, i.e. if there exists a stable probability distribution for the cube's side length x, in other words, *if ρ(x) exists*. Obviously, to perform the asked calculation this function needs to be specified; in which case there is no paradox.

An important lesson that is once more highlighted by Bertrand's paradox is that probability can only exist for variables (as edge length) that represent outcomes of *real physical events or systems*. According to our DEF1+2, probability is only defined for parameters that exhibit frequency stabilization *as can be shown by repeated physical tests including repetition of well-defined initial and probing conditions*. The corrected version of Bertrand's problem does contain the information needed to repeat the experiment (or do the calculation representing that experiment); in particular it states the *initial conditions* for doing these experiments. It contains the necessary information to 1) fabricate cubes with the specific (initiating) process of the factory (mathematically represented by the distribution density $\rho(x)$), and 2) to probe the object, namely by measuring its length, allowing to derive the proportion of lengths ≤ 1cm (one of the outcomes x). If $\rho(x)$ would not be given, but if one could perform the experiment with real machines, one would first have to verify whether the requested proportion stabilizes (which it will for say more than 100 cubes, and for almost all fabrication processes, except if there is a drift in process output parameters); in which case the requested probability is identified. But Bertrand's paradox obviously is posed as a mathematical question, therefore $\rho(x)$ has to be given. From the axioms of probability calculus and the definition of $\rho(x)$, it follows that the answer is given by $\int_0^1 \rho(x).dx$ (which would, for instance, be equal to ½ if $\rho(x)$ is Gaussian with mean value 1).

---

[13] It would indeed never be submitted to a student in mathematics or engineering in the mentioned form - except in alas nonexistent institutions where students would have to produce some foundational thinking. An acceptable formulation (because solvable) is: "A precision tool factory produces iron cubes with edge length x ≤ 2 cm, *and with edge length distribution density ρ(x) = ...* Etc."

In sum, according to our model, any mathematical formulation as Bertrand's paradox that is experimentally ambiguous, can safely be regarded as not well posed. More specifically, Bertrand's paradox can be seen as arising because the problem does not specify the *initial conditions* of the experimental trials that define the probability in question. The original version of Bertrand's paradoxes has exactly the same flaw. It goes as follows: "A chord is drawn randomly in a circle. What is the probability that it is shorter than the side of the inscribed equilateral triangle ?" (from [13] p.2). Bertrand showed that three different answers can be given, depending on exactly how the initializing randomization is interpreted. Indeed, there are many ways to 'randomly draw a chord' (which may not be obvious upon first reading of the problem). One can for instance randomly chose two points (homogeneously distributed) on the circle[14]; a procedure that leads to the probability 1/3, as can be measured and calculated. But other randomization procedures are possible (see [13] for a detailed treatment), leading in general, as already shown by Bertrand, to different outcomes. That Bertrand's paradoxes are not well posed is indeed the conclusion at which recent analyses arrive [3][15], as is obvious within our model.

R3. A popular idea is that, somehow, "probability depends on our knowledge, or on the subject". This is the key idea of subjective interpretations of probability, or (subjective) Bayesianism, associating probability with strength of belief. When I throw a regular die, I consider the probability for any particular throw to show a six to be 1/6. But what about my friend who is equipped with a sophisticated camera allowing him to capture an image of the die just before it comes to a halt ? For him the probability seems to be close to 0 or 1. Or what about following case: imagine Bob, waiting for a bus, only knowing that there is one bus passing per hour. He might think that the probability he will catch a bus in the next five minutes is low (say 5/60). Alice, who sits in a tower having a look-out over the whole city, seems to have a much better idea of the probability in case (she might quickly calculate, based on her observations, that it is close to 1). Are these not patent examples of the wide-spread idea that a same event can be given different probabilities, depending on the knowledge (or degree of belief) of the subject ? And is in that case probability not simply a measure of the strength of belief of the subject who attributes the probability ?

---

[14] for instance by fixing a spinner at the center of the circle; a pointer on the spinner and two independent spins generate two such independent random points.

[15] The latter reference also reviews von Mises' and Gnedenko's treatment of the problem, which will be seen to very well comply with ours ([13] p. 23). Keynes, for unclear reasons, holds fast to the principle of indifference (p. 23).

Paradoxical examples as these are unlimited, but if one adheres to our model, it is clear they stem from a neglect of the 'boundary conditions' that are part of the definition of probability. Probability is only defined – only exists – for repeatable experiments on well-defined systems, including, among others, well-defined conditions of observation. Doing a normal die throw, and observing the result on a table as is usually done, corresponds to a well-defined ρ-system: one can precisely describe the initiating system, the probing system etc. In the example, the second observer does not measure the same probability: he does not measure the probability of finding a six on a table after regular throwing, but of finding a six after measurement of whether a six will land or not on the table. The latter is a very different, and indeed fully deterministic, experiment; *at any rate, the observing subsystem is very different*. A similar remark holds for the bus-case; the measurement system (even if just the human eye) is part of the ρ-system; one cannot compare observer Alice and observer Bob if their means of observation are widely different. Again, only an experimental situation that is well-defined, by a specification of the probing subsystem, can have a defined probability.

If one can generalize examples as these, it seems that our definitions allow to safeguard the objective nature of probability (as a matter of fact, we believe that the just illustrated neglect of the probing subsystem is at the basis of subjectivist interpretations of one or the other form). If one includes in the experimental series all boundary conditions, and especially the probing subsystem, the probability of a given outcome *is* an objective measure. True, 'objective' (and 'observer-independent' even more) is a somewhat tricky word here: it means '*identical (and mind-independent) for all observers performing identical experiments*', so *objective in the scientific sense* – even if the observer is in a sense part of the system ! (Remember that the observing subsystem is part of the ρ-system.) Stated more precisely, the probability of an event is an 'objective' property of the ρ-system that generates the event, in that independent observers can determine or measure the same probability for the same event. Experience has massively shown that such well-defined experiments do lead to 'objective' probabilities: scientists all over the world attribute exactly the same probability to scientifically well-defined events (the paradigmatic case are the natural sciences, say quantum physics); and gamblers all over the world are confronted with always the same probabilities, much to the benefit of casino owners.

Our analysis might point out why probability lends itself easily to subjectivist interpretations of some kind (it seems easy to forget that in order to determine a probability of 'something', that something should include an observing system or conditions). Let us summarize our position as follows: according to our model probability is objective; to call it 'observer-independent' is true if understood well; the same holds for the idea that probability is an ontological, not epistemic category.

At this point the step to quantum mechanics is immediate. Indeed, it seems that our investigation has brought in focus a striking similarity between classical and quantum systems. It belongs to the key elements of the standard or Copenhagen interpretation that "the observer belongs to the quantum system", or at least that the measurement detectors belong to it, and influence it. Suffices here to cite Bohr in his famous debate with Einstein on quantum reality. As is well known, the issue at stake in the debate was the completeness of quantum mechanics, questioned by Einstein, Podolsky and Rosen in the case of two non-commuting observables of "EPR electrons". The key point of Bohr's counterargument is summarized in his phrase: "The procedure of measurement has an essential influence on the conditions on which the very definition of the physical quantities in question rests" [9]. According to Bohr and the Copenhagen interpretation, *the definition of quantum properties depends in a fundamental way on the measurement conditions*. Now quantum systems are probabilistic systems. Therefore the importance of the observation subsystem does not surprise us anymore. Indeed, we believe a careful inspection of the concept of probability shows that in *all* probabilistic systems, quantum or classical, the measurement system plays an essential role. We thus come to the surprising conclusion that, in this respect, quantum systems are not as exceptional as generally thought. Unfortunately we cannot elaborate here on this intriguing result, but it would be particularly interesting to apply it to quantum riddles such as Bell's theorem [14-16].

## 5. Conclusion.

We have provided in the present article an analysis of the concept of probability that can be seen as a variant of von Mises' frequency interpretation. Whether all essential aspects of what (objective) 'probability' is are captured by our model remains a question; but we believe that the uniqueness of Kolmogorov's axiomatic system, and the massive existence of scientific data to which the calculus applies, are convincing indications of the idea that one overarching objective interpretation is at least a defendable working hypothesis.

We proposed a model in which it only makes sense to define probability for systems or events that exhibit frequency stabilization, or p-randomness – the essential characteristic of probabilistic systems. It appeared useful to partition probabilistic systems into three parts: if natural, in object and environment; if artificial, in object, initializing subsystem, and probing subsystem. That a precise definition of initial and final conditions or subsystems is necessary for defining probability, is often neglected, and leads to countless paradoxes. Most importantly, including the *probing subsystem* into the probabilistic system allows to define probability in an objective, mind-independent manner – and thus to see probability as an objective, ontological category. *Only if the probing is defined, the probability is*. By the same token we showed that there is an essential parallel between quantum and classical systems: in order to be able to define a probability for an event, *in both cases* one needs to specify the 'observer', or rather the probing subsystem. The probability of an event depends on how the event is probed, whether the event is quantum or classical. Including the *initializing subsystem* into the probabilistic system, also allows to solve paradoxes, such as Bertrand's paradox.

Based on these ideas, we could conclude that probability does not exist for an object simpliciter, but for a *repeatable $\rho$-system*, or in other words for repeatable, $\rho$-identical events. Such $\rho$-systems involve objects under precise initial and final conditions. We could not define the problematic word 'identical' or 'similar' of some traditional definitions of probability better than in an operational manner, leading to the notion of '$\rho$-identical' events or systems. In this context, events and systems, and the objects and subsystems composing them, are identical or similar *as far as they lead to a probability density $\rho$*.

As a matter of fact, the notion of partitioning, or the idea that probability is only defined if all subsystems are defined, proved so useful (almost all our claims are based on it), that we summarized it in a slogan: *probability is a property of composed systems*. A somewhat more detailed phrasing is that probability is a property of certain repeatable and composed events (experiments).

Acknowledgements. We would like to acknowledge valuable comments of participants at the Conference of the Canadian Society for the History and Philosophy of Science in Montreal. Special thanks go to Mario Bunge and Yvon Gauthier for detailed discussion of the issues presented here.

______________________________________________

(*) louisvervoort@hotmail.com, louis.vervoort@umontreal.ca